# Vacuum Balloon – a 350-Year-Old Dream

Andrey Akhmeteli[*]
*LTASolid, Inc., 10616 Meadowglen Ln 2708, Houston, Texas 77042, USA*

Andrew V. Gavrilin [†]
*National High Magnetic Field Laboratory, 1800 E. Paul Dirac Dr., Tallahassee, Florida 32310, USA*

**The centuries-old idea of a lighter-than-air vacuum balloon has not materialized yet as such structure needs to be both light enough to float in the air and strong enough to withstand atmospheric pressure. We propose a design of a rigid spherical sandwich shell and demonstrate that it can satisfy these stringent conditions with commercially available materials, such as boron carbide ceramic and aluminum alloy honeycomb. A finite element analysis was employed to demonstrate that buckling can be prevented in the proposed structure. Also discussed are other modes of failure and approach to manufacturing.**

## Nomenclature

| | | |
|---|---|---|
| $E$ | = | compressive modulus of elasticity of the material of a thin homogeneous shell |
| $E_f$ | = | compressive modulus of elasticity of the face skin material of a sandwich shell |
| $G_3$ | = | honeycomb shear modulus |
| $h$ | = | thickness of a thin homogeneous shell |
| $h_1$ | = | thickness of the outer face skin of a sandwich shell, $h_1 = h_2$ |
| $h'_1$ | = | relative thickness of the outer face skin of a sandwich shell, $h'_1 = h_1/R$ |
| $h'_{1opt}$ | = | optimum relative thickness (analytical estimate) of the outer face skin of a sandwich shell |
| $h_2$ | = | thickness of the inner face skin of a sandwich shell, $h_2 = h_1$ |
| $h_3$ | = | thickness of the honeycomb core of a sandwich shell |
| $h'_3$ | = | relative thickness of the honeycomb core of a sandwich shell, $h'_3 = h_3/R$ |
| $h'_{3opt}$ | = | optimum relative thickness (analytical estimate) of the honeycomb core of a sandwich shell |

---

[*] e-mail: akhmeteli@ltasolid.com

[†] e-mail: gavrilin@magnet.fsu.edu

| | | |
|---|---|---|
| $K$ | = | a factor in a formula for face skin wrinkling analysis, $K = 0.95$ |
| $k_2$ | = | a factor in a formula for face skin wrinkling analysis, $k_2 = 0.82$ |
| $n$ | = | an exponent in a formula for intracell buckling analysis |
| $N$ | = | actual force per unit length of a sandwich plate |
| $N_{all}$ | = | allowable force per unit length of a sandwich plate (for shear crimping analysis) |
| $p_a$ | = | atmospheric pressure, $101\ kPa$ |
| $p_{cr}$ | = | critical buckling pressure |
| $q$ | = | payload fraction of a vacuum balloon |
| $R$ | = | outer radius of a shell |
| $S$ | = | honeycomb core cell size |
| $\lambda_{min}$ | = | minimum eigenvalue obtained in the finite element eigenvalue buckling analysis |
| $\mu$ | = | Poisson's ratio of a thin homogeneous shell |
| $\mu_f$ | = | Poisson's ratio of the face skins of a sandwich shell |
| $\rho_3$ | = | density of the honeycomb core of a sandwich shell |
| $\rho_3'$ | = | relative density of the honeycomb core of a sandwich shell, $\rho_3' = \rho_3/\rho_a$ |
| $\rho_a$ | = | atmospheric air density, $1.29\ kg \cdot m^{-3}$ |
| $\rho_f$ | = | density of the face skins of a sandwich shell |
| $\rho_f'$ | = | relative density of the face skins of a sandwich shell, $\rho_f' = \rho_f/\rho_a$ |
| $\rho_s$ | = | density of a thin homogeneous shell |
| $\sigma$ | = | compressive stress in a thin homogeneous shell |
| $\sigma_{dp}$ | = | critical uniaxial stress for intracell buckling |
| $\sigma_f$ | = | compressive stress in the face skins of a sandwich shell |
| $\sigma_{wr}$ | = | allowable uniaxial wrinkling stress (for face skin wrinkling analysis) |
| $\sigma_x, \sigma_y$ | = | stresses in the face skins in two orthogonal directions, $\sigma_x = \sigma_y = \sigma_f$ |
| $\tau_{xy}$ | = | shear stress in the face skins, $\tau_{xy} = 0$ |

# I. Introduction

THE idea of a lighter-than-air vacuum balloon is centuries old. In 1670, F. Lana di Terzi proposed a design of an airship where buoyancy was to be created by evacuated copper spheres (Ref. [1], see also Ref. [2] containing historical information related to the design). However, this dream has not materialized so far, because it is very difficult to design and manufacture a shell that is light enough to float in the air and strong enough to reliably withstand the atmospheric pressure. For example, A.F. Zahm (Ref. [3]) calculated the stress in a thin homogeneous one-layer rigid shell with vacuum inside and zero buoyancy, so that its mass equals that of the displaced air:

$$\frac{4}{3}\pi R^3 \rho_a = 4\pi R^2 h \rho_s, \quad (1)$$

where $R$ is the radius of the shell, $h$ is the shell thickness, $\rho_a$ and $\rho_s$ are the densities of air and of the shell material, respectively (we use an approximation for a thin shell). Let us then consider the condition of equilibrium for half of the shell (see Fig.1):

$$2\pi R h \sigma = \pi R^2 p_a, \quad (2)$$

where $\sigma$ is the compressive stress in the shell and $p_a = 101\ kPa$ is the atmospheric pressure at 0°C (we used a condition of equilibrium for a hemisphere of air in the atmosphere to calculate the right-hand side).

We obtain:

$$\frac{h}{R} = \frac{\rho_a}{3\rho_s}, \sigma = \frac{3}{2}\frac{\rho_s}{\rho_a}p_a. \quad (3)$$

If $\rho_s = 2700 kg \cdot m^{-3}$ (the density of aluminum), $\rho_a = 1.29 kg \cdot m^{-3}$ (the density of air at 0°C and 1 atm (101 kPa)), and $p_a = 1.01 \cdot 10^5 Pa$, then $\frac{h}{R} \approx 1.6 \cdot 10^{-4}, \sigma \approx 320\ MPa$, i.e., the stress is of the same order of magnitude as the compressive strength of contemporary aluminum alloys. It is important to note that this result does not depend on the radius of the shell.

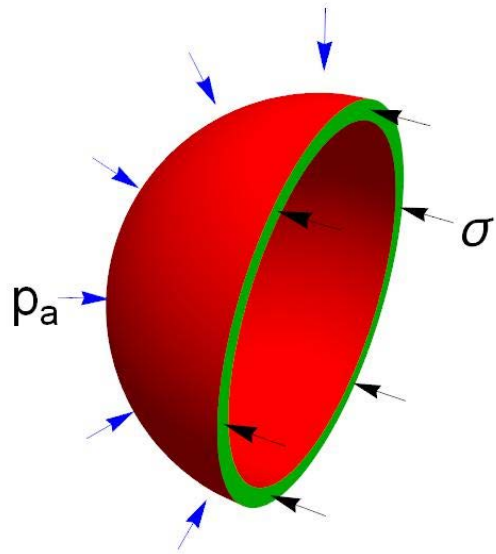

**Fig. 1 Stress and atmospheric pressure acting on a half of evacuated spherical shell (not to scale).**

A.F. Zahm notes that, while the results of stress calculation are quite problematic, buckling is an even more dangerous mode of failure for such a structure. Let us perform a simple buckling analysis for this structure (Ref. [4]). The critical buckling pressure for a thin spherical shell is given by the well-known formula of the linear theory of stability (Ref. [5]):

$$p_{cr} = \frac{2Eh^2}{\sqrt{3(1-\mu^2)}R^2}, \qquad (4)$$

where $E$ and $\mu$ are the modulus of elasticity and the Poisson's ratio of the material of the shell, respectively. If $p_{cr} = p_a$ and, e.g., $\mu = 0.3$, then

$$\frac{E}{\rho_s^2} = \frac{9p_a\sqrt{3(1-\mu^2)}}{2\rho_a^2} \approx 4.5 \cdot 10^5 kg^{-1}m^5s^{-2}. \qquad (5)$$

Even if we use diamond as the shell material ($E = 1.2 \cdot 10^{12} Pa$ and $\rho_s = 3500 kg \cdot m^3$), we obtain

$$\frac{E}{\rho_s^2} \approx 10^5 kg^{-1} m^5 s^{-2}. \tag{6}$$

In other words, even the maximally optimized homogeneous diamond shell of ideally spherical shape would inevitably fail already at $\sim 0.2 p_a$. Thus, one-layer shells made of any solid material in existence either cannot float in the air or have no chance of withstanding the atmospheric pressure. (It should be noted that we only considered the spherical shape in our analysis, as this shape is certainly the optimal one.)

Problems of this kind are quite common in aircraft design, and typical solutions are multi-layer shells with light core or stiffened shells.

In our patent application (Ref. [4]), we defined viable designs of a vacuum balloon based on three-layer shells made of commercially available materials. Numerous patents and articles on vacuum balloons had been published earlier (see, e.g., Ref. [6]), but, to the best of our knowledge, none of them properly addressed the crucial issue of buckling. More recently, other work addressing the issue of buckling for vacuum balloons was published (see, e.g., Refs. [7-11] and references there). The recent popular articles on vacuum balloons (Refs. [12-13]) also reflect current interest in this topic.

As our design (Ref. [4]) garnered some interest, it is advisable to describe it here, in a journal article format, after significant rework, providing details of the all-important buckling analysis. Detailed comparison with the vastly different designs featured in related work by others (Refs. [7-11]) would perhaps require finite-element analysis of each proposed structure and is beyond the scope of this article. We would just like to note that our design, unlike many others, is spherically symmetric and scalable (multiplying all linear dimensions by the same factor provides an equally viable design; see some caveats related to intracell buckling in Section III), so it has fewer parameters (as there is no dependence on the polar and azimuthal angles or absolute linear dimensions), which facilitates its analytical optimization. It is also noteworthy that the design does not contain any components under tension. This may be advantageous for using materials such as ceramics, whose compressive properties are typically much better than the tensile properties.

## II. A Sandwich Vacuum Balloon and Its Buckling Analysis

As an example, let us consider a three-layer spherical shell with face skins of equal thickness $h_1 = h_2$ and a core of aluminum alloy honeycomb of thickness $h_3$ (see Fig. 2 and Fig. 3 below).

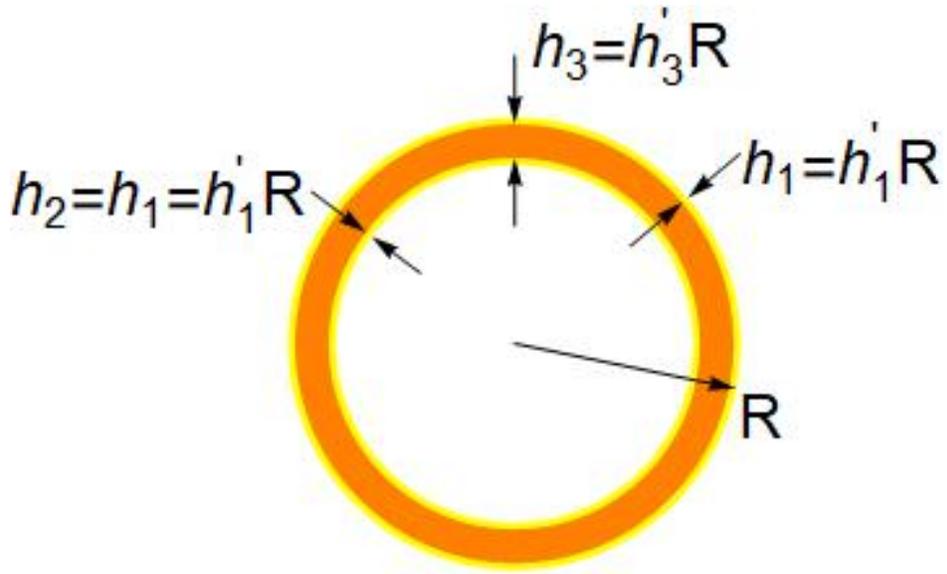

**Fig 2 Dimensions of a spherical sandwich shell (not to scale).**

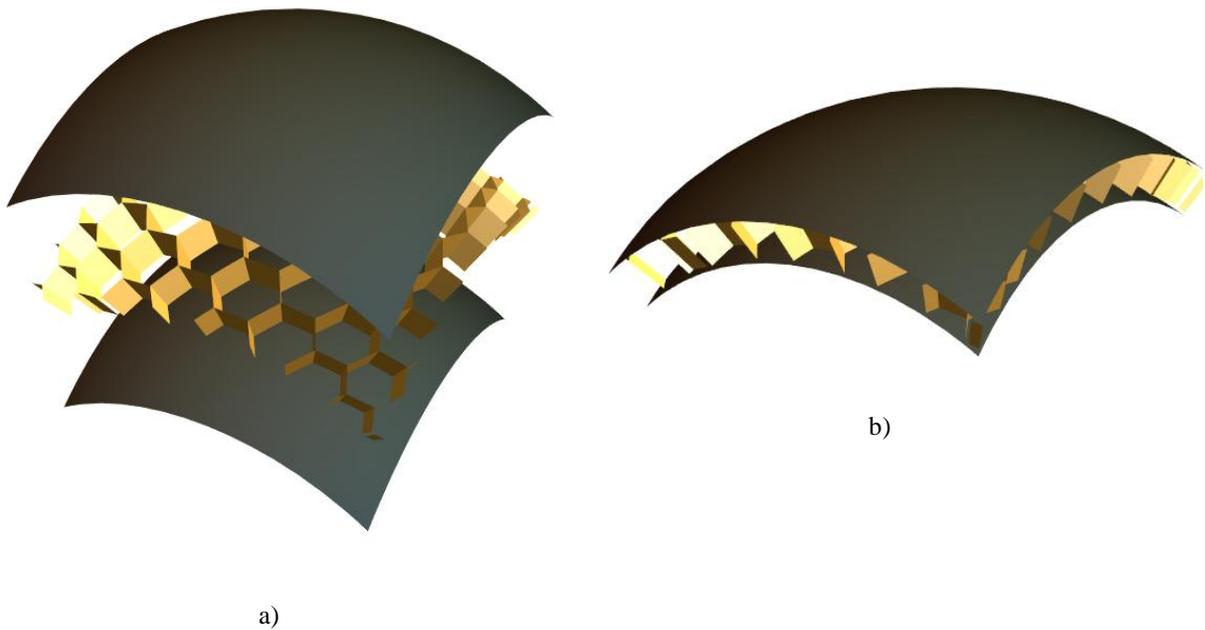

**Fig 3 A fragment of a spherical sandwich shell a) before and b) after assembly (not to scale).**

In order to prove the design feasibility, we used parameters of commercially available materials in our study. Boron carbide ceramic was chosen as the face skin material (density $\rho_f = 2500\ kg \cdot m^{-3}$, elasticity modulus $E_f = 460\ GPa$, compressive strength $\sigma_f = 3200\ MPa$, Poisson's ratio $\mu_f = 0.17$ (Ref. [14])). PLASCORE PAMG-XR1-3.1-1/8-7-N-5056 honeycomb was chosen as the core material (cell size 1/8 inch (3.2 $mm$), nominal foil gauge 0.0007 inch (18 $\mu m$), nominal density 3.1 pcf (50 $kg \cdot m^{-3}$), bare compressive strength 340 psi (2.3 $MPa$) / modulus 97 ksi (670 $MPa$), plate shear strength 250 psi (1.7 $MPa$) ("L"), 155 psi (1.1 $MPa$) ("W") / modulus 45 ksi (310 $MPa$) ("L"), 20 ksi (140 $MPa$) ("W") (Ref. [15])).

If $R$ is the radius of the shell, we assume that $R \gg h_3 \gg h_1$. To assess the design feasibility, we also anticipate that the shell allows a small payload fraction $q = 0.1$ (the ratio of the mass of the payload at zero buoyancy and the mass of the displaced air). Then, the condition of zero buoyancy has the following form:

$$\tfrac{4}{3}\pi R^3 \rho_a (1-q) = 4\pi R^2 \left(2h_1 \rho_f + h_3 \rho_3\right), \tag{7}$$

or

$$6h_1' \rho_f' + 3h_3' \rho_3' = 1 - q, \tag{8}$$

where $h_1' = \frac{h_1}{R}, h_3' = \frac{h_3}{R}, \rho_f' = \frac{\rho_f}{\rho_a}, \rho_3' = \frac{\rho_3}{\rho_a}$.

The buckling stability condition that we used is described by the following semi-empirical formula for critical pressure obtained for three-layer domes (Ref. [16]):

$$p_{cr} = 2E_f \frac{h_1(h_3+h_1)}{R^2} \approx 2E_f \frac{h_1 h_3}{R^2} = 2E_f h_1' h_3'. \tag{9}$$

In this case, $E_f$ is the modulus of elasticity of the face skin material, and $p_{cr}$ is the maximum pressure at which the three-layer shell is stable. The requirements for core rigidity are discussed below, but let us first find the values of $h_1'$ and $h_3'$ that maximize $p_{cr}$. Using Eq. (8), let us eliminate $h_1'$ from Eq. (9):

$$p_{cr} = 2E_f \frac{1-q-3h'_3\rho'_3}{6\rho'_f} h'_3. \tag{10}$$

The value of $p_{cr}$ is maximal for

$$h'_3 = h'_{3opt} = \frac{1-q}{6\rho'_3} \approx 3.9 \cdot 10^{-3}. \tag{11}$$

In that case the optimal values of $h'_1$ and $p_{cr}$ are:

$$h'_1 = h'_{1opt} = \frac{1-q-3h'_3\rho'_3}{6\rho'_f} \approx 3.9 \cdot 10^{-5}, p_{cr} \approx 1.37\, p_a. \tag{12}$$

This is a good indication that the design is feasible. However, we need to assess the buckling stability more accurately and consider other modes of failure. We cannot rely on the results of the above analytical approach as it hinges on the semi-empirical formula Eq. (9) from Ref. [16]. The validity limits of this formula are not clear, in particular, it is not clear how this formula should be modified to take into account manufacturing imperfections when they are different from those in the shells of Ref. [16]. The results of the above approach were verified and optimized by a finite element analysis (FEA) using ANSYS, which enabled us to compute the stress and strain in the shell components and to perform the eigenvalue buckling analysis (which is actually a classical Euler buckling analysis). The results of the FEA analysis confirmed that the analytical approach provides reasonable estimates and a good starting point for optimization in our case. However, we base the conclusions of this article on the results of the FEA analysis, not on the results of the analytical approach.

For the FEA analysis, a 2D axisymmetric model in the spherical system of coordinates, with due regard for corresponding boundary conditions at the edges, was found sufficient and adequate (Fig. 4). In this model, PLANE82 2D high-order 8-node elements, which are well suited for curved boundaries, were used for the finite element mesh in axisymmetric mode. The mesh was heavily refined, and the elements' aspect ratios were appropriately adjusted.

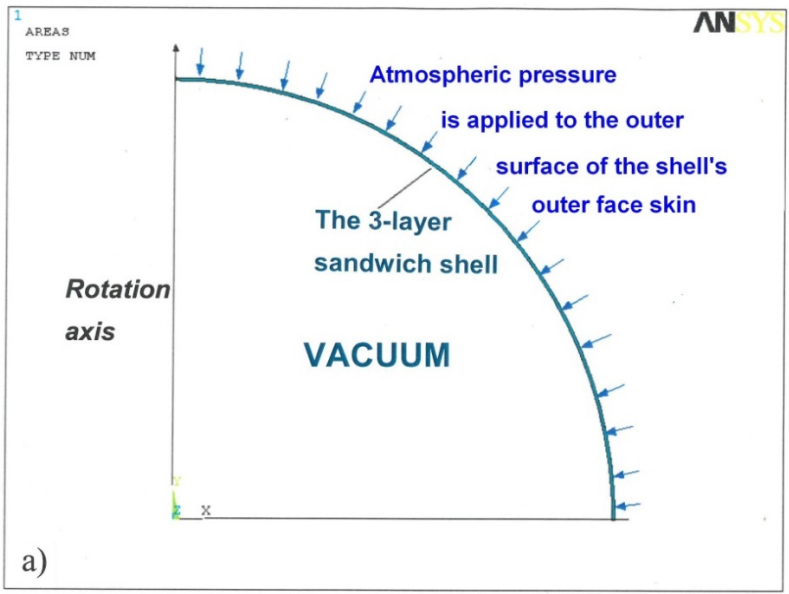

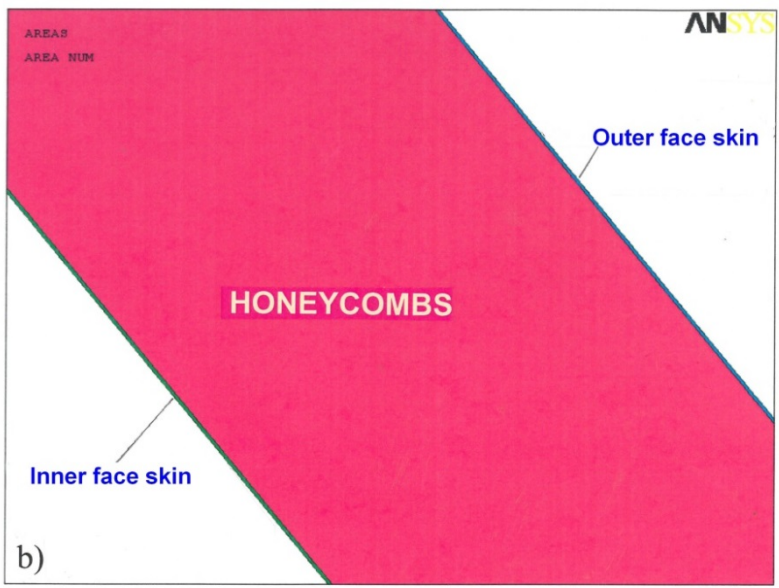

**Fig. 4  The ANSYS FEA: (a) a 2D axisymmetric model and (b) an enlarged fragment of the 3-layer sandwich shell's solid model.**

In the FEA, the anisotropic material properties of the honeycomb were treated in accordance with recommendations of a honeycomb manufacturer (Ref. [17], p, 20). For the sake of simplicity, we assume that the honeycomb is a transversely isotropic material, so the lesser of the two values of shear modulus from Ref. [15] was used (which makes the results more conservative). Thus, we used the following values: the out-of-plane component of the modulus – $670\ MPa$; the in-plane components of the modulus – $67\ Pa$; out-of-plane components of the shear modulus – $140\ MPa$; the in-plane component of the shear modulus – $14\ Pa$; the Poisson's ratio – $10^{-5}$ (very small values of the Poisson's ratio and the in-plane components of the modulus and the shear modulus were used to avoid singularities, in accordance with Ref. [17], p, 20).

Thus, the ANSYS eigenvalue buckling analysis input includes the loads. The output of the analysis is the eigenvalues (buckling load multipliers), which are the safety factors for buckling modes (for the input loads). The minimum eigenvalue $\lambda_{min}$ obtained in the eigenvalue buckling analysis determines the critical buckling load. As we load our structure with atmospheric pressure $p_a$, we have the following relation between the minimum eigenvalue $\lambda_{min}$ and the critical buckling pressure:

$$\lambda_{min} = \frac{p_{cr}}{p_a}. \tag{13}$$

The optimized parameters of the analytical approach were used as initial values for optimization through the FEA. The eigenvalue $\lambda_{min}$, regarded as a function of $h'_3$, has a rather sharp maximum of $\lambda_{min} \approx 2.65$ (see Fig. 5) for a value of $h'_3 \approx 3.53 \cdot 10^{-3}$, which is close to the value we arrived at using the simplified method. The corresponding value of $h'_1$ approximates $4.23 \cdot 10^{-5}$. To give an idea of how the eigenvalue $\lambda_{min}$ varies with the payload fraction $q$, let us note that $\lambda_{min}$ is approximately 3.21 for an optimized design with zero payload fraction.

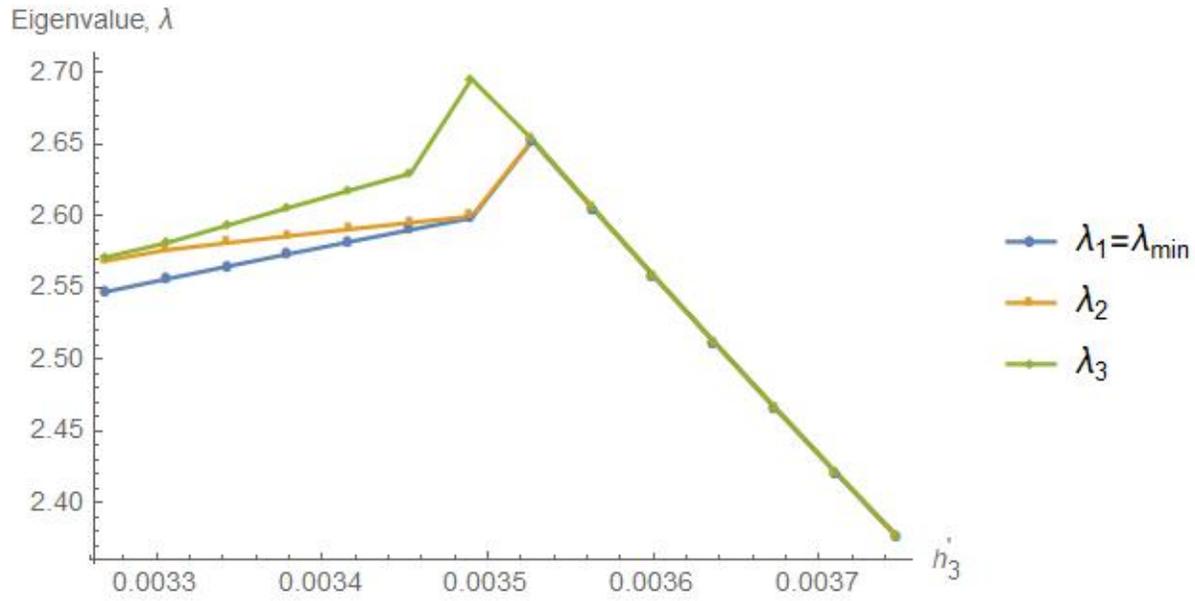

**Fig. 5 The 3 least eigenvalues $\lambda_{min}$, $\lambda_2$, and $\lambda_3$ from the ANSYS eigenvalue buckling analysis versus the relative core thickness $h'_3$ for payload fraction $q = 0.1$.**

The safety factor of 2.65 is not very high, as empirical knock-down factors are typically applied to the results of small-deflection analysis for externally-pressurized thin-walled spherical shells to take into account initial imperfections and other factors. For example, the knock-down factor of 0.2 is recommended in Ref. [18, p. 6-9] for hemispherical sandwich domes (there is a very good agreement between our FEA results and the results obtained with the use of formulas in Ref. [18, p. 6-6] with a knock-down factor of 1 for buckling critical pressure/stress). The formulas of Ref. [18] are based on the solution from Refs. [19, 20]). If we perform linear (and, if required, non-linear) buckling analysis with due regard for imperfections of manufacturing, the safety factors will decrease, but the obtained results are high enough to reasonably expect that the safety factors will still be quite sufficient for the state-of-the-art manufacturing accuracy, as thoroughly manufactured thin spherical shells were shown to withstand external pressure of up to 80-90% of the critical one (Refs. [21, 22]). The relevant variation of thickness of the shallow spherical shells in the experiments of Refs. [21, 22] was about ±1% of the thickness. In our design, if the radius is $2.5\ m$, the thickness of the sandwich shell is about $9\ mm$ (see the dimensions in Conclusion), so the comparable manufacturing accuracy would be about $\pm 0.1\ mm$. While such accuracy may be difficult to achieve, the knock-down coefficient of 0.2 is conservative. According to the review of experimental results on buckling of spherical shells in Ref. [23], the knock-down factors for various technologies often exceed 0.4, which would be sufficient in our case (for payload factor of

0.1). If necessary, incoming testing of sandwich plates can be performed before the assembly of the shell to make sure the knock-down factor for the plates exceeds the required value.

Only homogeneous spherical shells are discussed in Refs. [21-23], and information on spherical sandwich shells is scarce, but the results and recommendations of Ref. [16] suggest similar conclusions.

Taking into account the imperfections in the finite-element analysis is beyond the scope of this work as the imperfections depend on the specific technologies.

### III. Other Modes of Failure

Now let us verify that other modes of failure are not problematic for the design. We used standard unity checks from Refs. [17, 24]. Providing detailed descriptions of the failure modes and explanations of the standard formulas here does not seem warranted.

Let us first check that the compressive stress $\sigma_f$ in the face skins does not exceed the compressive strength of the face skin material. Instead of Eq. (2) we have:

$$2\pi R \cdot 2h_1 \sigma_f = \pi R^2 p_a, \quad \sigma_f = \frac{p_a}{4h_1'} \approx 600 \, MPa, \quad (14)$$

which is much less than the compressive strength of boron carbide (3.2 $GPa$).

The following formula is used to check the design for shear crimping (Ref. [17, pp. 8, 19], Ref. [24, pp. 245-246]):

$$N_{all} = h_3 G_3 \approx 0.49 \, MPa \cdot R, \quad (15)$$

where $N_{all}$ is the allowable force per unit length of a sandwich plate in some direction and $G_3 = 20 \, ksi \approx 138 \, MPa$ is the honeycomb shear modulus. The actual force per unit length is much less:

$$N = 2h_1 \sigma_f \approx 51 \, kPa \cdot R, \quad (16)$$

The design was checked for face skin wrinkling using the following formula (Ref. [24, p. 234]):

$$\sigma_{wr} = k_2 E_f \sqrt{\frac{E_3 h_1}{E_f h_3}}, \tag{17}$$

where $\sigma_{wr}$ is the allowable uniaxial wrinkling stress, $k_2 = 0.82$ is a theoretically derived coefficient for honeycomb cores, $E_f = 460\ GPa$ is the face skin modulus, $E_3 = 97$ ksi $\approx 0.69$ GPa is the honeycomb modulus. We obtain: $\sigma_{wr} \approx 1.58\ GPa$. However, we have biaxial stress, so we should check that (Ref. [24, p. 235])

$$\frac{(\sigma_x^3 + \sigma_y^3)^{\frac{1}{3}}}{K\sigma_{wr}} + \left(\frac{\tau_{xy}}{\sigma_{wr}}\right)^2 < 1, \tag{18}$$

where $\sigma_x$ and $\sigma_y$ are stresses in two orthogonal directions ($\sigma_x = \sigma_y = \sigma_f \approx 600\ MPa$), $\tau_{xy}$ is the shear stress ($\tau_{xy} = 0$), $K = 0.95$. The left side of Eq. (18) equals approximately 0.5, so this check also yields a satisfactory result.

The design is scalable with respect to all of the above modes of failure: an equally successful design can be obtained by multiplying all linear dimensions by the same factor. However, this is not true for another mode of failure – so called intracell buckling (also known as dimpling). We use the following formula (Ref. [24], p. 241):

$$\sigma_{dp} = \frac{2E_f}{1-\mu_f^2}\left(\frac{h_1}{S}\right)^2, \tag{19}$$

where $\sigma_{dp}$ is the critical stress for intracell buckling and $\mu_f = 0.17$ is the Poisson's ratio of the face skins, and $S = \frac{1}{8}\ inch$ (3.2 mm) is the cell size. We obtain $\sigma_{dp} \approx 168\ MPa \cdot m^{-2} \cdot R^2$. However, we have biaxial stress, so we must make sure that (Ref. [24, p. 242])

$$\frac{(\sigma_x^n + \sigma_y^n)^{\frac{1}{n}}}{\sigma_{dp}} + \left(\frac{\tau_{xy}}{0.8\ \sigma_{dp}}\right)^2 < 1, \tag{20}$$

where $n = 3$ if the cell size $S > 15.63\ h_1$ (that means $R < 4.8\ m$ for our values of $S$ and $h_1'$) and $n \geq 3$ otherwise (see Ref. [24], p. 243). If $R \geq 4.8\ m$, we have

$$\sigma_f < 2^{-\frac{1}{3}}\sigma_{dp} \leq 2^{-\frac{1}{n}}\sigma_{dp}, \qquad (21)$$

and condition (20) is satisfied. If $R < 4.8\ m$, we have $\sigma_f < 2^{-\frac{1}{3}}\sigma_{dp} \approx 134\ MPa \cdot m^{-2} \cdot R^2$, or $R > 2.11\ m$. Thus, we obtain the following condition of stability for intracell buckling: $R > 2.11\ m$.

We did not study possible effects of small leaks in the face skins, but they should not present a greater problem than for other vacuum systems, as only rough vacuum is required for vacuum balloons. If pressure difference in neighboring honeycomb cells is a concern, one may need to use a honeycomb with cell perforations (Ref. [15]).

## IV. Towards a Prototype Vacuum Balloon

Manufacturing of the boron carbide face skins seems to be the most challenging part of the design, as they may be very thin, and their density needs to be close to the theoretical boron carbide density, otherwise the elasticity modulus can be insufficient.

For large prototypes ($R \geq 25\ m$), the face skin thickness exceeds $1\ mm$, and parts of the face skins can be produced using traditional methods, such as uniaxial pressing with subsequent sintering (Refs. [25, 26]).

For smaller prototypes, the thickness of the face skins is $0.1\ mm$ by order of magnitude. Producing such parts is technologically challenging, and the parts may be too fragile. Detailed treatment of these issues is beyond the scope of this article, but a preliminary discussion is clearly necessary.

The face skins can be produced either by deposition on a sacrificial substrate (this can be time-consuming if the process is to yield high elasticity modulus) or using gelcasting, which can provide "fine features down to $100\ \mu m$ scale" (Ref. [27]). Another approach to manufacturing uniform spherical boron carbide shells with a thickness of the order of $100\ \mu$ by dropping a slurry coating on a molybdenum substrate and subsequent drainage and curing is described in Ref. [28]. While the radius of the shells in Ref. [28] is small ($1\ mm$), the method was used with different materials to manufacture shells of a radius of up to $375\ mm$ (Ref. [29]). Shells of larger radius can probably be manufactured by varying the viscosity of the slurry and/or using rotation.

To circumvent the issue of fragility, one can first bond the face skins to the honeycomb and then remove the face skin supports (the substrates or parts of the molds).

To fabricate the entire vacuum balloon, one will need to join several sandwich panels using some standard approach, such as bonded butt joints using H sections (Ref. [30], p. 5). The weight penalty is not estimated in this article, but it will be smaller for larger sandwich panels.

According to Ref. [31], "in terms of volume efficiency, …convex stellated shells are comparable to spherical shells with a knockdown factor of 0.65", so the former can achieve better buckling efficiency than the latter "when the effects of geometric imperfections are considered". If a similar conclusion is also true for sandwich shells, then sandwich stellated shells can be another option for vacuum balloons. In terms of manufacturing, such shells can be attractive as they can be assembled using flat sandwich panels.

The exterior and the interior of the structure can be connected with a valve at a modest weight penalty for initial evacuation of the structure or for altitude control.

The buoyant force reduction due to the shell compression by atmospheric pressure was calculated to be less than 0.4%.

Point loading of the structure should be avoided. For example, it is relatively easy to provide distributed support with larger contact area for the structure on the ground.

We did not discuss issues related to adhesives here (mass requirements, modes of failure, etc.), but these issues are less significant for shells of larger radius, as the adhesive mass scales as $R^2$, and the mass of the entire structure scales as $R^3$. Neither did we discuss potential use of more exotic materials (such as chemical vapor deposition (CVD) diamond for the face skins or architected cellular materials (Ref. [13]) for the core) to significantly increase safety factors and/or the payload fraction.

## V. Conclusion

We showed that a lighter-than-air rigid vacuum balloon can theoretically be built using commercially available materials. The design of this article is an evacuated spherical sandwich shell of outer radius $R > 2.11\ m$ containing two boron carbide face skins of thickness $4.23 \cdot 10^{-5} \cdot R$ each that are reliably bonded to an aluminum honeycomb core of thickness $3.52 \cdot 10^{-3} \cdot R$. The structure is lighter than air (it allows a payload fraction of 0.1) and can withstand the atmospheric pressure. For example, if $R = 2.5\ m$, the face skin thickness is $106\ \mu m$, the honeycomb thickness is $8.8\ mm$, the mass of the shell is $75.7\ kg$, the payload capacity for zero buoyancy is $8.7\ kg$.

A prototype vacuum balloon would also become the first ever lighter-than-air solid (for example, aerogels are actually not lighter than air due to the air inside).

Manufacturing a prototype vacuum balloon will be a major breakthrough. It will require detailed engineering to resolve numerous relatively minor issues, but the results of this article suggest that we can finally bring to fruition the ancient dream of a vacuum balloon.

It took mathematicians 357 years to prove Fermat's Last Theorem. Will it take us more to build the first vacuum balloon?

## Acknowledgments

A portion of this work is performed at the National High Magnetic Field Laboratory, which is supported by the National Science Foundation Cooperative Agreement DMR 1644779 and the State of Florida.

The authors are grateful to A. N. Palazotto for his interest in this work and valuable remarks. One of the authors (A. A.) is grateful to A. I. Afanasyev, who initiated this work by an inquiry on applications of high-strength materials. The authors are grateful to anonymous reviewers for valuable remarks.

## References

[1] Lana, F., *Prodromo. Ouero saggio di alcune inuentioni nuoue premesso all'Arte Maestra*, Rizzardi, Brescia, 1670, Chap. 6 (in Italian),
https://books.google.ru/books?id=o7AGGIKz0_wC&pg=PP9&hl=ru&source=gbs_selected_pages&cad=2#v=onepage&q&f=false, accessed 07/08/2018

[2] Shikhovtsev, E., http://mir.k156.ru/aeroplan/de_bausset_aeroplane-03-1.html#a03-1-16, (in Russian and in English), accessed 02/05/2019

[3] Zahm, A. F., *Aërial Navigation: A Popular Treatise on the Growth of Air Craft and on Aeronautical Meteorology*, D. Appleton and Company, New York and London, 1911, p. 443, https://books.google.com/books?id=hRdDAAAAIAAJ&pg=PA443&lpg=PA443&dq=%22zahm%22+vacuum+balloon&source=bl&ots=HO8PwEw0M5&sig=AQGKWimBz3oa9Y32TwTnEgAu-UQ&hl=en&sa=X&ved=2ahUKEwiFroi-s4LfAhUq9YMKHWFHBtwQ6AEwAXoECAkQAQ#v=onepage&q=%22zahm%22%20vacuum%20balloon&f=false, accessed 12/02/2018

[4] Akhmeteli, A. M., Gavrilin, A. V., U.S. Patent Application for "Layered shell vacuum balloons," Appl. No. 11/517915, filed 08 Sep. 2006.

[5] Timoshenko, S. P., Gere, J. M., *Theory of Elastic Stability,* McGraw Hill, New York, 1961, p. 512.

[6] Armstrong, L. M., Peoria, IL, U.S. Patent 1390745 for an "Aircraft of the lighter-than-air type," patented 13 Sep. 1921.

[7] Barton, S. A., Florida State University Research Foundation (Tallahassee, FL, US), U.S. Patent 7708161 for "Light-weight vacuum chamber and applications thereof," filed 05 Dec. 2006, issued 04 May 2010.

[8] Snyder, J. W., Palazotto, A. N., Finite Element Design and Modal Analysis of a Hexakis Icosahedron Frame for Use in a Vacuum Lighter-Than-Air Vehicle, *Journal of Engineering Mechanics*, 2018, **144**, (6), pp 04018042.

[9] Adorno-Rodriguez, R., Palazotto, A. N., Nonlinear Structural Analysis of an Icosahedron Under an Internal Vacuum, *Journal of Aircraft*, May–June 2015, **52**, (3), pp 878-883.

[10] Rapport, N., Middleton, W. I., U.S. Patent Application for "Lighter-Than-Air Fractal Tensegrity Structures," Appl. No. 14/807118, filed 23 July. 2015.

[11] Jenett, B. E., Gregg, C. E., Cheung, K. C., Discrete Lattice Material Vacuum Airship, *AIAA SciTech Forum, 7-11 January 2019, San Diego, California*, pp 1-12.

[12] Ball, P., Flying on empty, *New Scientist*, 21/28 December 2019, pp 68-69.

[13] Surcouf, O., Dirigeables: le miracle du vide, *Science&Vie*, No 1233, June 2020, pp 90-93 (in French), https://www.science-et-vie.com/technos-et-futur/dirigeables-le-miracle-du-vide-56281, accessed 7/29/2020.

[14] http://www.skylinecomponents.com/B4C.html, accessed 12/25/2018.

[15] https://www.plascore.com/honeycomb/honeycomb-cores/aluminum/pamg-xr1-5056-aluminum-honeycomb-core/, accessed 12/25/2018.

[16] Brix, G., Durchschlagen von GFP-Sandwichkuppeln bei gleichförmigem Außendruck, *IfL-Mitt. (350. Mitteilung aus dem Institut für Leichtbau und ökonomische Verwendung von Werkstoffen, Dresden*), 1968, **7,** (11), pp 408-413 (in German).

[17] https://www.hexcel.com/user_area/content_media/raw/Honeycomb_Sandwich_Design_Technology.pdf, accessed 12/25/2018.

[18] SULLINS, R. T., SMITH, G. W. and SPIER, E. E., Manual for Structural Stability Analysis of Sandwich Plates and Shells, NASA CR-1457, 1969, https://apps.dtic.mil/dtic/tr/fulltext/u2/a310684.pdf, accessed 8/1/2020.

[19] YAO, J. C., Buckling of Sandwich Sphere Under Normal Pressure, *Journal of the Aerospace Sciences*, March 1962, pp 264-268.

[20] PLANTEMA, F. J., *Sandwich Construction: The Bending and Buckling of Sandwich Beams, Plates and Shells,* John Wiley & Sons, Inc., New York/London/Sydney, 1966.

[21] KRENZKE, M. A. and KIERNAN, T. J., Elastic Stability of Near-Perfect Shallow Spherical Shells, *AIAA Journal*, 1963, **1**, (12), pp 2855-2857.

[22] KRENZKE, M. A. and KIERNAN, T. J., Erratum: Elastic Stability of Near-Perfect Shallow Spherical Shells, 1964, *AIAA Journal*, **2**, (4), pp 784.

[23] BŁAŻEJEWSKI, P., MARCINOWSKI, J., ROTTER, M., Buckling of externally pressurised spherical shells. Experimental results compared with recent design recommendations, *ce/papers*, 2017, **1**, (2&3), pp 1010-1018.

[24] COLLIER, C., Consistent Structural Integrity and Efficient Certification with Analysis, Vol.3, AFRL-VA-WP-TR-2005-3035, 2005, https://apps.dtic.mil/dtic/tr/fulltext/u2/a444085.pdf, accessed 12/27/2018.

[25] KAISER, A., Hydraulic Pressing of Advanced Ceramics, *cfi/Berichte der DKG*, 2007, **84**, No. 6, pp E 27-32, http://www.alpha-ceramics.de/system/00/01/52/15245/633855139353281250_1.pdf, accessed 8/10/2020.

[26] KAISER, A., LUTZ, R., Uniaxial Hydraulic Pressing as Shaping Technology for Advanced Ceramic Products of Larger Size, *Interceram*, 2011, No. 03-04, pp 230-234, http://www.laeis.eu/System/00/01/95/19513/634559894557055155_1.pdf, accessed 8/10/2020.

[27] LU, R., CHANDRASEKARAN, S., DU FRANE, W. L., LANDINGHAM, R. L., WORSLEY, M. A., KUNTZ, J. D., Complex shaped boron carbides from negative additive manufacturing, *Materials and Design*, 15 June 2018, **148**, pp 8-16.

[28] RUICHONG CHEN, JIANQI QI, LIN SU, QIWU SHI, XIAOFENG GUO, DI WU, TIECHENG LU, ZHIJUN LIAO, Rapid preparation and uniformity control of B$_4$C ceramic double curvature shells: Aim to advance its applications as ICF capsules, *Journal of Alloys and Compounds*, 2018, **762**, pp 67-72.

[29] LEE, A., BRUN, P.-T. , MARTHELOT, J. , BALESTRA, G. , GALLAIRE, F. , REIS, P.M., Fabrication of slender elastic shells by the coating of curved surfaces, *Nature Comm.*, 2016, **7:11115**, pp 1-7, https://www.nature.com/articles/ncomms11155.pdf, accessed 8/17/2020.

[30] Sandwich Panel Fabrication Technology, Hexcel LTU 018, 2001, https://studylib.net/doc/18103284/sandwich-panel-fabrication-technology, accessed 7/21/2019.

[31] XIN NING, PELLEGRINO, S., Searching for imperfection insensitive externally pressurized near-spherical thin shells, *Journal of the Mechanics and Physics of Solids*, November 2018, **120**, pp 49-67.